\documentclass[proceedings, preprint]{rmaa}

\usepackage{paralist}
\usepackage{psfrag,color}
\usepackage{wrapfig}
\usepackage{multicol}
\usepackage{fixltx2e}

\SetYear{2008}
\SetConfTitle{A LONG WALK THROUGH ASTRONOMY}

\title{SEARCHING FOR COMPACT STAR CLUSTERS IN M81 USING HST/ACS IMAGES} 

\author{
  M. Santiago-Cort\'es,\altaffilmark{1} 
  Y. D. Mayya,\altaffilmark{1}
  D. Rosa-Gonz\'alez,\altaffilmark{1}}

\altaffiltext{1}{Instituto Nacional de Astrof\'isica \'Optica y Electr\'onica, Apartado Postal 51 y 216, Puebla, M\'exico (scortes@inaoep.mx, ydm@inaoep.mx, danrosa@inaoep.mx).}

\shortauthor{SANTIAGO-CORT\'ES, MAYYA \& ROSA-GONZ\'ALEZ}
\shorttitle{SEARCHING FOR COMPACT STAR CLUSTERS IN M81}

\listofauthors{M. SANTIAGO-CORT\'ES, Y. D. MAYYA, D. ROSA-GONZ\'ALEZ}
\indexauthor{SANTIAGO-CORT\'ES, M.}
\indexauthor{MAYYA, Y. D.}
\indexauthor{ROSA-GONZ\'ALEZ, D.}

\abstract{We study the stellar cluster population in M81, using the \textit{HST/ACS} images in the filters F435W, F606W and F814W that cover a total field of view of $\sim140\,arcmin^{2}$. We present details about the selection criteria, which were based both on morphological and photometrical features. The extracted sample of stellar clusters shows the presence of two cluster populations, a blue cluster group (young) with 560 objects, and a red cluster group (old) with 120 objects. The young group lacks clusters more massive than $10^{4}\,M_{\odot}$, that are present in large numbers in its neighbor M82. The luminosity distribution function of the young group follows a power-law distribution with $\alpha = 2.0$, whereas that for the red group resembles very much the globular cluster luminosity function in the Milky Way. Assuming an age of 5 Gyr, these red clusters have masses between $10^{5}$ and $10^{7}\,M_{\odot}$.}

\resumen{
La poblaci\'on de c\'umulos estelares en M81 es estudiada utilizando observaciones del \textit{HST/ACS} en los filtros F435W, F606W y F814W, las cuales cubren un \'area de $\sim140\,arcmin^{2}$. Presentamos detalles de los criterios de selecci\'on de los candidatos a c\'umulos estelares compactos, el cual est\'a basado en caracter\'isticas morfol\'ogicas y fotom\'etricas. En la muestra extraida de c\'umulos estelares tenemos la presencia de dos poblaciones de c\'umulos. Una poblaci\'on de c\'umulos azules (j\'ovenes) con 560 objetos y una poblaci\'on de c\'umulos rojos (viejos) de 120 objetos. La poblaci\'on de c\'umulos j\'ovenes carece de c\'umulos estelares m\'as masivos que $10^{4}\,M_{\odot}$, que est\'an presentes en gran n\'umero en su vecina M82. La distribuci\'on de la funci\'on de luminosidad para los c\'umulos j\'ovenes es consistente con una ley de potencias con un \'indice $\alpha =2.0$, mientras que la funci\'on de luminosidad para los c\'umulos rojos sigue la misma distribuci\'on que los c\'umulos globulares de la V\'ia L\'actea. Asumiendo una edad de $5\times10^{9}$ a\~nos estos c\'umulos tienen masas entre $10^{5}$ y $10^{7}\,M_{\odot}$.}

\addkeyword{Galaxies: Individual (M81)}
\addkeyword{Galaxies: star clusters}
\addkeyword{Galaxies: catalogs}

\begin{document}
\maketitle

\section{INTRODUCTION}
\label{sec:intro}
With the advent of the \textit{HST (Hubble Space Telescope)} a new class of clusters have been identified: the Compact Star Clusters (CSCs) with typical masses of $\sim3\times10^{4}$ to $10^{6}\,M_{\odot}$ and sizes between 1 and 6 pc \citep{Meurer1995}. CSCs have been found in several environments, including violent star forming regions within interacting galaxies \citep{Whitmore1999}. M81 (NGC 3031) is a large Sab spiral galaxy at 3.63 Mpc  \citep{Freedman1994} and it belongs to the M81 Group, which includes the prototype starburst galaxy M82. Almost all the stars in the disk of M82 were formed in a violent disk-wide burst about 100--500 Myr ago, following the interaction of M82 with M81 \citep{Mayya2006}. Recent observations of M82 show that it has a large number of CSCs, with young clusters (age $<10$ Myr) concentrated towards the center and old clusters ($\sim100$ Myr) homogeneously distributed across the disk. The cluster population of M81 has been studied previously by \citet{Chandar2001} using data from the \textit{WFPC2/HST} camera discovered 114 CSCs. They observed an area of $40\,arcmin^{2}$ in the disc of M81. The analysis found two different cluster populations, red clusters which are candidate for globular clusters and a young population with ages ($< 600$ Myr) that support a recent star formation epoch in M81 related with the interaction between M81 and M82. In this analysis we extend their work by analyzing a new set of data with improved spatial resolution and covering an area which is 3.7 times larger than the area covered by \citet{Chandar2001}. The resulting sample of CSCs is the biggest for M81 to date.

\section{OBSERVATIONS}
\label{sec:obser}
The observations used in this work were taken with the \textit{ACS} on \textit{HST}. There are 12 fields that cover a field of view of $\sim140\,arcmin^{2}$, with a resolution of $0.05^{''}pix^{-1}$  (0.88 pc $pix^{-1}$). The F435W (B), F606W (R) and  F814W (I) filters were used on each field to carry out the cluster search (Figure 1).

\begin{figure}[h]
  \includegraphics[width=\columnwidth]{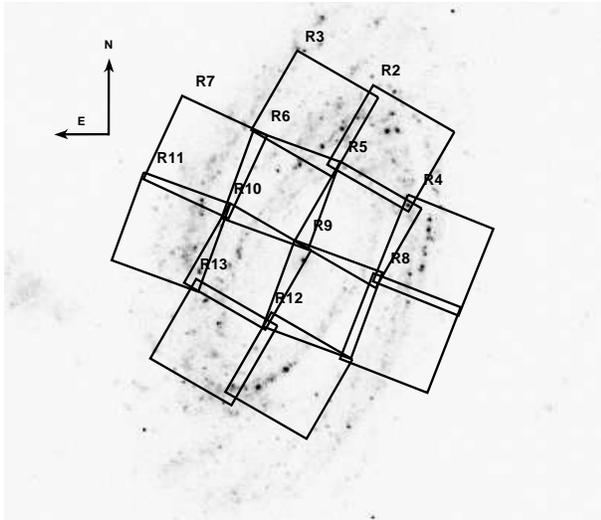}
  \caption{A $23'\times20'$ \textit{GALEX} image of M81 with the \textit{HST/ACS} fields used in this work plotted and labeled.}
  \label{fig:simple}
\end{figure}

\section{CLUSTER DETECTION  AND SELECTION}
We used the automatic detection code SExtractor \citep{Bertin1996} to create an unbiased sample of cluster candidates. The B band was used for the first selection of candidates, and we carried out photometry in each of the B, R and I images. This resulted in 261,844 objects in the 12 fields analyzed. The preliminary list of candidates contains both unresolved (stellar-like) and resolved (extended) objects. Among the resolved objects, we have the CSCs, and also two kinds of contaminating sources: 1) sources formed by improper subtraction of the background due to the presence of dust and complex small-scale disk structures and 2) sources formed by the superposition of several point sources mainly due to stellar crowding. In order to reject these contaminating sources we applied the following criteria:
Selection Filter 1:\\
A) We considered  all those sources with $2.4 < FWHM < 10$ pixels and area greater than 50 pix as cluster candidates. Almost all the sources with FWHM less than 2.4 pixels are point sources (stellar-like), whereas objects with FWHM $> 10$ do not satisfy the compactness criteria. Extended objects such as large stellar groups and artificial sources created by residual background have area less than 50 pix, and are rejected by our selection criteria (Figure 2).

\begin{figure}[h]
    \centering\vskip -4mm
    \includegraphics[width=\columnwidth]{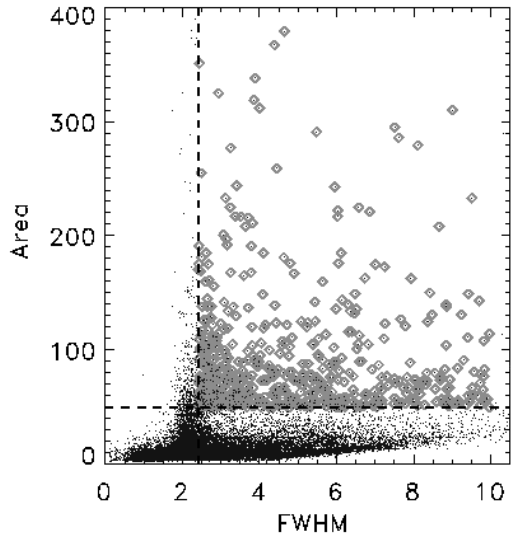}
    \vskip -5mm
    \caption{Stellar-like sources and extended sources are included in the sources detected by SExtractor (points). The compact stellar cluster candidates are those with FWHM $> 2.4$ pix and Area $> 50$ pix (rhombus).}
 \label{fig:2}
\end{figure}

Selection Filter 2:\\
A) All cluster candidates with ellipticity $< 0.1$ are compatible with being bound systems and are automatically accepted. B) Majority of the sources with ellipticities greater than 0.1 are in general not clusters. However, we found that SExtractor measures high ellipticities for some genuine sources when they are surrounded by diffuse background or multiple stars. These genuine sources are centrally peaked and hence could be rescued by comparing aperture magnitudes estimated in different radii. We found that if the difference between the  aperture magnitudes of diameters 2 and 4 pixels is less than 1.5 mag, then they are true clusters. Notice that at fainter magnitudes artificial sources due to crowding can still enter our sample even with the application of these filters. An extra selection criterion based on colors and magnitudes was used in the analysis stage to filter out these sources. After using the two different filter selection criteria described above, we detected 680 compact stellar cluster candidates, 244 of which are brighter than B=22 mag. Having in mind that \citet{Chandar2001} detected 114 clusters reaching at V=22 mag, ours is the largest sample reported in M81 until now.

\section{ANALYSIS}
In Figure 3, we present the color histogram distribution for the clusters candidates. They have a range in color between $-0.23 < B - I < 4.0$ and magnitudes between $17.8<B< 24.3$ mag (solid line). It can be seen at brighter magnitudes ($17.8<B<23.0$, dotted line) that the clusters appear to divide themselves into two groups at $B-I=1.7$ (arrow), although this division is less pronounced at fainter magnitudes (dashed line). This division was found previously by \citet{Chandar2001}. Based in this trend we separated the cluster sample into two groups: a blue group with $B-I < 1.7$ and a red group with $B-I > 1.7$ (see also Figure 4, dashed-point line).

\begin{figure}[h]
    \centering
    \includegraphics[width=\columnwidth]{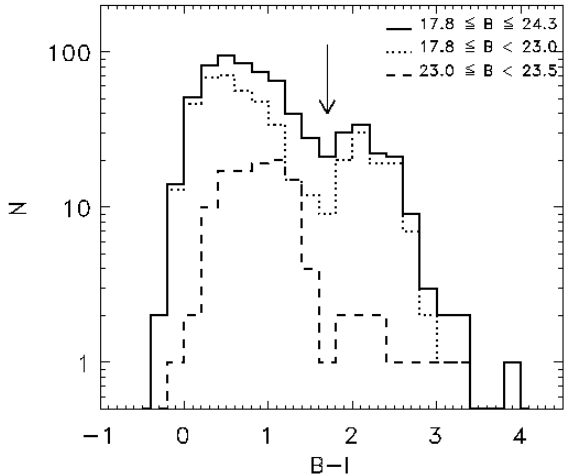}
    \vskip -3mm
    \caption{Color histogram distribution for CSCs population are plotted in different range of magnitudes. The arrow marks the color ($B-I=1.7$) that divide the sample in two populations observed until magnitude $B=23.0$.}
 \label{fig:3}
\end{figure}

In Figure 4, we present the Color-Magnitude Diagram (CMD) of the selected candidates. They are divided into two families at $B-I=1.7$. There are 560 stellar clusters in the blue group with $B-I < 1.7$ (blue rhombus) and 120 in the red group with $B-I > 1.7$ (red circles). The reddening vector for a visual extinction of 1 mag is shown \citep{Cardelli1989}. It is very unlikely that the two families are produced by just the effect of dust. An evolutionary track for a Simple Stellar Population (SSP, \citet{Girardi2002}) for a cluster mass of $10^{4}\,M_{\odot}$ is shown superposed on the points for the blue group. It can be easily seen that there are no clusters above the brightest point in the track, which indicates that there are no massive blue clusters in our sample. The track reproduces the observed colors of the brightest clusters ($B<22$ mag). At fainter magnitudes there are several clusters bluer than the SSP which share the same color range as that for individual stars. Many of these sources may not be clusters and instead blended stars and hence we exclude them from detailed analysis. Therefore, most genuine clusters are consistent with ages younger than 500 Myr.

The red clusters have colors expected for galactic globular clusters. We hence assumed a constant age of 5 Gyr for this group, and estimated their masses. The resulting masses are in the range of $10^{5}-10^{7}\,M_{\odot}$. Notice that a change in the age of factor of 2 implies a change in the mass by a factor of 2 in this range of ages.

For the young clusters the luminosity function (LF, Figure 5) in the magnitude range $19<B<22$ mag is consistent with a power law index of $\alpha = 2.0$ (dotted line), similar to that found for the young stellar cluster systems in other galaxies \citep{Elson1985}. At fainter magnitudes the LF is affected by incompleteness and the contamination by bright stars. It can be noticed that the number of detected clusters starts to decrease from the expected numbers with no clusters detected brighter than $B=18$ mag. This indicates the absence of clusters more massive than $10^{4}M_{\odot}$, which are present in large numbers in its neighbor M82.

The LF for the old group clusters is plotted in Figure 6 along with that for the Milky Way globular clusters from \citet{Harris1996}.  It can be seen that the two distributions agree very well for $M_{B} < -4$ mag $(B=24 mag)$, which is the completeness limit of our observations. Notice that, we have detected 120 globular clusters, which compares well with the 146 globular clusters in the Milky Way. It is heartening to find that the numbers and the LF between M81 and the Milky Way coincide, in spite of the complexity involved in the selection process.

\begin{figure*}[t]
    \centering
    \includegraphics[width=1.45\columnwidth]{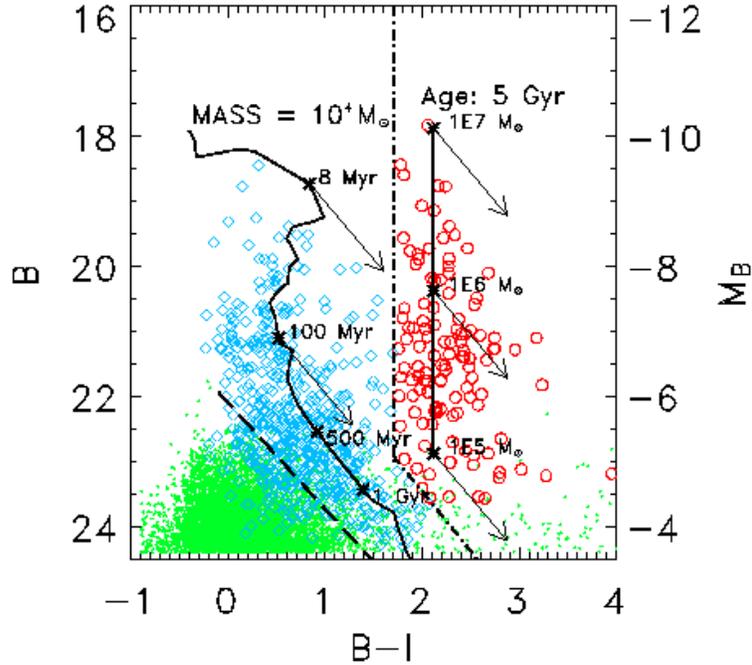}
    \vskip -4mm
    \caption{CMD for the M81 CSCs population: blue rhombus for blue group and red circles for red group divided by the dashed-point line. The small green points in the bottom part of the figure represent the position of unresolved sources which are mainly field stars in M81. An evolutionary track for an SSP \citep{Girardi2002} of mass$=10^{4}\,M_{\odot}$ and age ranging from 8 Myr to 1 Gyr is shown. Colors of blue group CSCs suggest ages less than 500 Myr when they are brighter than $B=22$ mag. Fainter clusters are severely affected by stellar contamination. An SSP of constant age of 5 Gyr, but with masses in the range of $10^{5}$--$10^{7}\,M_{\odot}$ is also shown. It fits the colors and magnitudes for the red group which are globular cluster candidates. Reddening vectors corresponding to $Av=1$ mag are shown at selected positions on the tracks \citep{Cardelli1989}.}
\label{fig:4}
\end{figure*}

\begin{figure}[!h]
    \centering
    \includegraphics[width=0.85\columnwidth]{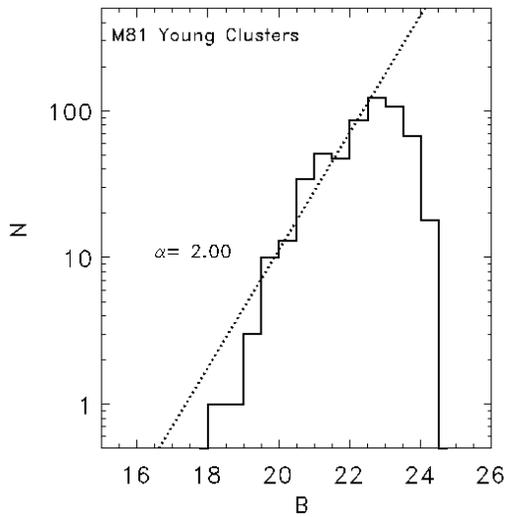}
    \vskip -3mm
    \caption{The LF of the young group (solid line). A power law of $\alpha = 2.0$ (dotted line) is also plotted.}
 \label{fig:5}
\end{figure}
\begin{figure}[!h]
    \centering
    \small
    \includegraphics[width=0.85\columnwidth]{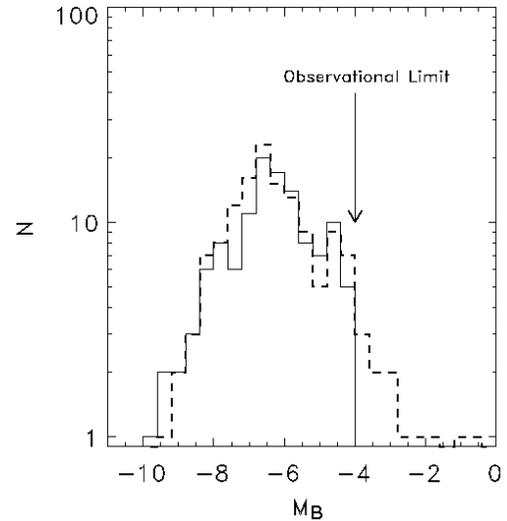}
    \vskip -3mm
    \caption{The LF of globular clusters detected in M81 (solid line) and the Milky Way (dashed line). The observational limit for extended sources is also marked.}
 \label{fig:6}
\end{figure}

\section{DISCUSSION AND CONCLUSIONS}
\label{sec:conclus}
Thanks to the superb spatial resolution of the \textit{ACS} camera on board the \textit{HST}, we were able to obtain the largest sample of CSCs in M81 until now. We found a total population of 680 CSCs. The sample is divided into two well-defined populations --- a blue group with ages less than 500 Myr and masses around $10^{4}\,M_{\odot}$ and a red group of old globular clusters with masses between $10^{5}$ and $10^{7}\,M_{\odot}$. It may be noted that encounter of M82 with M81 happened around 500 Myr ago which could have triggered the formation of blue clusters, as speculated by \citet{Chandar2001}. However we cannot rule out the presence of clusters older than 500 Myr, as these objects would lie below our completeness limit. Hence the observed population could as well be due to the normal star formation occurring in the disk of M81 independent of the interaction. The luminosity function of young clusters follows a power law distribution with $\alpha=2.0$, with a clear deficit of clusters more massive than $10^{4}\,M_{\odot}$ as compared to the number of massive stellar clusters observed in M82 \citep{Mayya2008}. Perhaps massive cluster formation requires an extreme burst mode of star formation such as seen in M82.

We find that the luminosity function for the M81 globular clusters has the same distribution as that for the Milky Way globular clusters with a peak at $M_{B}=-6.7$  mag ($M_{V}= -7.5 mag$). The result confirms the similarities between M81 and the Milky Way but also makes our complex selection method very trustworthy. The close encounter of M82 with M81 possibly created a new generation of compact clusters in the disk but did not affect the distribution of old clusters that were in place at the time of the encounter.

\section{ACKNOWLEDGMENTS}
We would like to thank the Hubble Heritage Team at the Space Telescope Science Institute for making the reduced fits files available to us. This work was supported by CONACyT (M\'exico) grants: fellowship number 4464 and project numbers 58956-F and 49942-F.

\end{document}